\documentclass[aps,prb,preprint,showpacs,superscriptaddress]{revtex4}
\usepackage{graphicx}
\usepackage{color}
\usepackage[cp1251]{inputenc}
\usepackage{amsmath}
\usepackage{amssymb}
\usepackage{color}
\usepackage{epsfig}
\usepackage{floatflt}
\usepackage{hyperref}

\def\c{ \cdot }

\def\p{ \partial }

\def\eq{\equiv}

\def\q{\quad}

\def\equ{Eq.~\eqref}

\begin{document}

\bibliographystyle{apsrev}

\title{Resonance effects due to the excitation of surface Josephson plasma waves
in layered superconductors}

\author{V.A.~Yampol'skii}
\affiliation{Advanced Science Institute, The Institute of Physical
and Chemical Research (RIKEN), Wako-shi, Saitama, 351-0198, Japan}
\affiliation{ A.Ya.~Usikov Institute for Radiophysics and
Electronics Ukrainian Academy of Sciences, 61085 Kharkov, Ukraine}

\author{A.V.~Kats}
\affiliation{ A.Ya.~Usikov Institute for Radiophysics and
Electronics Ukrainian Academy of Sciences, 61085 Kharkov, Ukraine}

\author{M.L.~Nesterov}
\affiliation{ A.Ya.~Usikov Institute for Radiophysics and
Electronics Ukrainian Academy of Sciences, 61085 Kharkov, Ukraine}

\author{A.Yu.~Nikitin}
\affiliation{ A.Ya.~Usikov Institute for Radiophysics and
Electronics Ukrainian Academy of Sciences, 61085 Kharkov, Ukraine}

\author{T.M.~Slipchenko}
\affiliation{ A.Ya.~Usikov Institute for Radiophysics and
Electronics Ukrainian Academy of Sciences, 61085 Kharkov, Ukraine}

\author{S.E.~Savel'ev }
\affiliation{Advanced Science Institute, The Institute of Physical
and Chemical Research (RIKEN), Wako-shi, Saitama, 351-0198, Japan}
\affiliation{Department of Physics, Loughborough University,
Loughborough LE11 3TU, UK}

\author{Franco Nori}
\affiliation{Advanced Science Institute, The Institute of Physical
and Chemical Research (RIKEN), Wako-shi, Saitama, 351-0198, Japan}
\affiliation{Department of Physics, Center for Theoretical
Physics, Applied Physics Program, Center for the Study of Complex
Systems, University of Michigan, Ann Arbor, MI 48109-1040, USA}

\begin{abstract}
We analytically examine the excitation of surface Josephson plasma
waves (SJPWs) in periodically-modulated layered superconductors.
We show that the absorption of the incident electromagnetic wave
can be substantially increased, for certain incident angles, due
to the resonance excitation of SJPWs. The absorption increase is
accompanied by the decrease of the specular reflection. Moreover,
we find the physical conditions guaranteeing the total absorption
(and total suppression of the specular reflection). These
conditions can be realized for Bi2212 superconductor films.

\end{abstract}

\maketitle

\section{Introduction}

Over the last twenty years, the physical properties of layered
superconductors have attracted the attention of many research
groups . The strongly-anisotropic high-temperature  $\rm
Bi_2Sr_2CaCu_2O_{8+\delta}$ single crystals are the most prominent
members of this family. Numerous experiments on the
$\mathbf{c}$-axis transport in layered high-$T_c$ superconductors
(HTS) justify the use of a model in which the superconducting
CuO$_2$ layers are coupled, through the block layers, by the
intrinsic Josephson effect. The Josephson current flowing along
the $\mathbf{c}$-axis is coupled with the electromagnetic field
inside the insulating dielectric layers, thereby causing a
specific kind of elementary excitations called Josephson plasma
waves (JPWs) (see, e.g., Ref.~\onlinecite{JPWs}). In other words,
the layered structure of Bi-based superconductors and related
compounds favors the propagation of electromagnetic waves through
the layers. These waves are of considerable interest because of
their terahertz (THz) and sub-THz frequency ranges, which are
still hardly reachable for both electronic and optical devices.
The frequencies of terahertz waves are in the region of resonance
frequencies of molecules and are expected to have many
applications.

The unusual optical properties of layered superconductors,
including their reflectivity and transmissivity, caused by the
JPWs excitation, were studied, e.g., in Ref.~\onlinecite{helm}.
Earlier works on this problem have focused on the propagation of
bulk waves, that is possible in the frequency range above the
Josephson plasma frequency $\omega_{_J}$, at $\omega>\omega_{_J}$,
only. The presence of the sample boundary can produce a new branch
of the wave spectrum below the Josephson plasma frequency,
$\omega<\omega_{_J}$, i.e., surface Josephson plasma waves
(SJPWs), which are an analog of the surface plasmon
polaritons~\cite{Plasman,Agrmils}. Recently, the existence of
SJPWs in layered superconductors in the THz frequency range was
predicted~\cite{yam1,sloika}. Surface waves play an important role
in many fundamental resonance optics phenomena~\cite{bliokh}, such
as the Wood's anomalies in reflectivity~\cite{Agrmils,reflect} and
transmissivity~\cite{transmis,KNN_05} of periodically-corrugated
metal samples. A recent overview of unusual resonators can be
found in Ref.~\onlinecite{bliokh}. Therefore, it is essential to
study similar resonance phenomena caused by the excitation of
surface waves in layered superconductors.

The dispersion curve, $\omega(q)$, of the surface waves lies below
the ``vacuum light line'', $\omega = c q$, where $q$ is the
wave-number and $c$ is the speed of light. This means that the
surface waves have wave-vectors greater than the wave-vectors of
light of the same frequency in vacuum. Thus, to excite the surface
waves by means of incident irradiation, it is necessary to use
special methods~\cite{Agrmils}, such as, e.g., the attenuated
total reflection (ATR) method and the surface modulation method.

In this paper, we study the excitation of surface Josephson plasma
waves while diffracting the electromagnetic wave incident onto the
periodically-modulated layered superconductor. For simplicity, we
present results for single-resonance cases, when only one SJPW is
excited.  The excitation of SJPWs affects the absorption and
reflection of the incident electromagnetic waves, specifically,
determining their resonance dependence on the frequency $\omega$
and the incident angle $\theta$. These phenomena are potentially
useful for detecting THz radiation.

\section{Model}

\begin{figure}
\begin{center}
\begin{center}
\vspace*{-5cm}
\includegraphics[width=17cm]{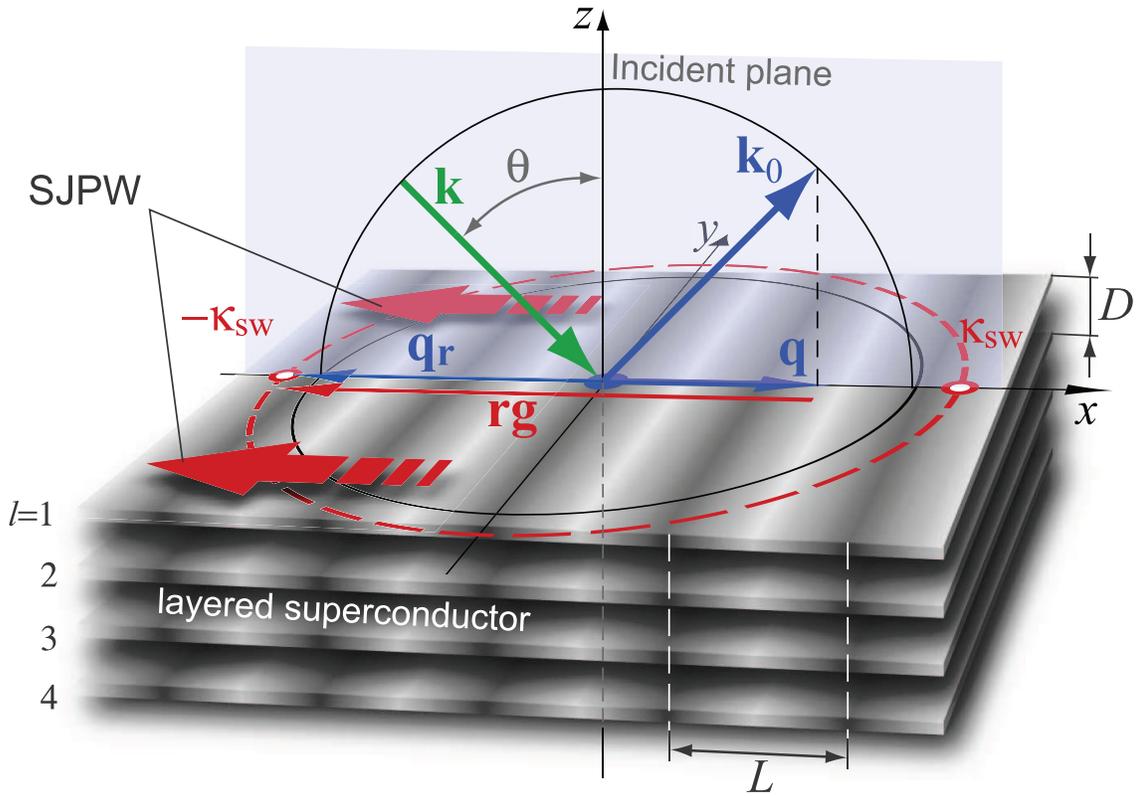}
\end{center}
\caption{(Color online) Geometry of the problem: ${\mathbf{k}}$
and ${\mathbf{k_0}}$ are the wave-vectors of the incident and
specularly reflected waves, $\kappa_{\rm sw}$ is the wave-number
of the SJPW. The case of backward resonance diffraction in the
$r$th order ($r <0$) is shown. Also,  $q_{r}=q+rg \simeq
-\kappa_{\rm sw}$ denotes the tangential component of the
wave-vector of the resonance wave, and $g=2\pi/L$ is the period of
the reciprocal grating. }\label{graph1}
\end{center}
\end{figure}

Consider a semi-infinite layered superconductor in the simplest
geometry shown in Fig.~\ref{graph1}. The crystallographic
$\mathbf{ab}$-plane coincides with the $xy$-plane and the
$\mathbf{c}$-axis is directed along the $z$-axis. Superconducting
layers are numbered by an integer $l\geq 1$.

Suppose that the maximum $\mathbf{c}$-axis Josephson current
density, $J_c$, is periodically modulated in the $x$-direction
with a spatial period $L$. Such a modulation can be realized, for
instance, either by irradiating a standard $\rm
Bi_2Sr_2CaCu_2O_{8+\delta}$ sample covered by a modulated
mask~\cite{kwok} or by pancake vortices controlled by an
out-of-plane magnetic field~\cite{koshelevprl}.

A $p$-polarized (transverse magnetic) plane monochromatic
electromagnetic wave with electric, $\mathbf{E}=\left\{E_x, 0,
E_z\right\}$, and magnetic, $\mathbf{H}=\left\{0, H, 0\right\}$,
fields is incident onto a periodically-modulated layered
superconductor at an angle $\theta$ from the vacuum half-space.
The in-plane and out-of-plane components of its wave-vector are
    $$
k_x \eq q = k \sin\theta, \q
k_z =-k\cos\theta , \q k=\omega/c .
    $$
The in-plane periodic modulation results in generating the
diffracted waves with in-plane and out-of-plane wave-vector
components,
    $$
q_n=q+ng , \q    k_{zn}^{V}=\sqrt{k^2-q_n^2} , \q \mathrm{Re}
[k_{zn}^{V}],\,\mathrm{Im} [k_{zn}^{V}] \ge 0 ,
    $$
$n$ is an integer and $g=2\pi/L$. The resonance excitation of the
SJPWs corresponds to the condition,
\begin{equation}\label{1}
 q_n = k\sin\theta+ng = {\rm sign}(n) \mathrm{Re} [\kappa_{\rm sw}(\omega)],
\end{equation}
where $\kappa_{\rm sw}(\omega)>k$ is the SJPW
wave-number~\cite{sloika},
\begin{equation}\label{25_12_07_1}
\kappa_{\rm sw}^2(\omega)  = k^2 \left[ 1 -
\displaystyle\frac{k^2\lambda_{ab}^2\Omega^2}{\varepsilon
(1-\Omega^2-i\nu\Omega)}\right]^{-1} .
\end{equation}

The total magnetic field in the vacuum ($z>0$) is given by the
Fourier-Floquet expansion,
\begin{eqnarray}\label{ashva}
  &H^V(x,z)=H^{\rm inc}\Bigl[\exp (iq x - i k z\cos\theta )+\nonumber \\
  &\sum\limits_{n} {R_{n} \exp (iq_{n}x + ik_{zn}^{V} z)}\Bigr],
\end{eqnarray}
where $H^{\rm inc}$ is the amplitude of the incident wave and $R_{n}$ are
the transformation coefficients (TCs). The time dependence
$\exp(-i\omega t)$ is omitted hereafter.

Using Maxwell equations, we express the tangential component of
the electric field in the vacuum in terms of the magnetic field,
\begin{eqnarray}\label{ashva4}
 &E_x^V(x,z) =H^{\rm inc}\Bigl[-\beta^{V}\exp (iq x - ik z \cos\theta) \nonumber \\
 &+ \sum\limits_{n} {\beta_{n}^{V}R_{n}\exp(iq_{n} x + ik_{zn}^{V}
z)}\Bigr],
\end{eqnarray}
where $\beta^{V}=\cos\theta$, $\beta_{n}^{V}=k_{zn}^{V}/k$.

The electromagnetic field within the layered superconductor is
related to the gauge-invariant phase difference, $\varphi_l$, of
the order parameter. The values of $\varphi_l$ in the junctions
are governed by the set of coupled sine-Gordon equations (see,
e.g., Ref.~\onlinecite{art}),
\begin{eqnarray}\label{0}
&\displaystyle\left( 1-\frac{\lambda_{ab}^2}{D^2}
\partial_l^2\right) \left( \frac{\partial ^2\varphi_l}{\partial
t^2}+\omega_r\frac{\partial
\varphi_l}{\partial t} + \omega_{_J}^2(x) \sin\varphi_l \right)\nonumber\\
&- \,\displaystyle\frac{c^2}{\varepsilon}\frac{\partial ^2
\varphi_l}{\partial x^2} =0.
\end{eqnarray}
Here $\lambda_{ab}$ is the London penetration depth across the
layers, $D$ is the spatial period of the layered structure, the
discrete second derivative operator $\partial_l^2$ is defined as
$\partial_l^2 f_l = f_{l+1}+f_{l-1}-2f_l$,
\[
\omega_r=\frac{4\pi\sigma_c}{\varepsilon}
\]
is the relaxation frequency, $\sigma_c$ is the quasi-particle
conductivity across the layers,
\begin{equation}\label{2}
\omega_{_J}(x) = \sqrt{\frac{8\pi e D J_c(x)}{\hbar\varepsilon}}
\end{equation}
is the periodically-modulated Josephson plasma frequency, and
$\varepsilon$ is the interlayer dielectric constant. The Fourier
expansion of $\omega_{_J}^{2}(x)$ is
\begin{equation}\label{omegap}
\omega_{_J}^{2}(x)=\omega_{_J}^{2}\left[1+\sum\limits_{n\neq0}
{f_{n}} \exp(i n g x)\right] , \q f_{-n} = f_{n}^{\ast}.
\end{equation}
Below we assume the modulation to be small, $|f_{n}| \ll 1$.

As was shown in Ref.~\onlinecite{kosh}, the intralayer
quasi-particle conductivity, $\sigma_{ab}$, should also be taken
into account if $\omega$ is far enough from the Josephson plasma
frequency. The contribution of the in-plane conductivity to the
dissipation can be easily incorporated in our analysis. However,
for the frequency range considered here (close to $\omega_{_J})$,
this contribution is strongly suppressed and can be safely omitted
because the relative value of the term with $\sigma_{ab}$ is
\[
\left(\frac{\lambda_{ab}}{\lambda_c}\right)^2
\left(\frac{\sigma_{ab}}{\sigma_{c}}\right)
\left|1-\frac{\omega}{\omega_{_J}}\right|\ll \,1.
\]
Here $\lambda_c = c/(\omega_{_J} \sqrt{\varepsilon}$) is the London
penetration depth along the layers.

For Josephson plasma waves, the nonlinear equations \eqref{0} can
be linearized, i.e., $\sin\varphi_{l}$ can be replaced by
$\varphi_{l}$. We also assume that the gauge-invariant phase
difference experiences small changes, $|\varphi_{l+1} -
\varphi_{l}| \ll |\varphi_{l}|$, and thus we can use the continuum
approach, replacing $D^{-1}\p_l \varphi_{l}$ by $\p_z \varphi(z)$.
Then \equ{0} yields
\begin{equation}\label{e88}
\displaystyle\left(1-\lambda_{ab}^2 \frac{{\partial ^2
}}{{\partial z^2 }} \right)\left(\omega\!_{_J}^2(x) -\omega^2
- i \omega \omega_r  \right)\varphi
- \displaystyle\frac{c^2}{\varepsilon}\frac{\partial^2
\varphi}{\partial x^2}=0.
\end{equation}

The magnetic and electric
fields are related to the gauge invariant phase difference as
\begin{equation}\label{e89}
\frac{{\partial \varphi}}{{\partial x}} =\frac{{2\pi D}}{{\Phi _0
}}\left(1- \lambda_{ab} ^2 \frac{{\partial ^2 }}{{\partial z^2 }} \right) H,
\end{equation}
\begin{equation}\label{granysl2}
E_x = -i k \lambda_{ab}^2 \frac{{\partial H}}{{\partial z}},\quad E_z=ik\frac{\Phi_0}{2\pi D}\varphi
\end{equation}
where $\Phi_0=\pi c \hbar/e$ is the flux quantum and $e$ is the elementary charge.

\section{Diffraction of the electromagnetic field}

Inside the layered superconductor, we represent the
gauge-invariant phase difference and the electromagnetic field as expansions
over the eigenfunctions,
\begin{equation}\label{26_02_08}
\varphi(x,z) = H^{\rm inc} \sum\limits_{s} \bar{C}_{s } \bar{\Psi}_{s}(x) \exp ( p_s
z),
\end{equation}
\begin{equation}\label{9_11_07}
H(x,z) = H^{\rm inc} \sum\limits_{s} C_{s } \Psi_{s}(x) \exp ( p_s
z),
\end{equation}
\begin{equation}\label{elsup}
E_x (x,z) =  - H^{\rm inc} \sum\limits_{s}  a _s C_s \Psi_{s}(x)
\exp (p_s z),
\end{equation}
with
\begin{equation}\label{magsup}
\bar{\Psi}_{s}(x) = \sum\limits_{n} \bar{\Psi} _{s|n} \exp (iq_{n} x ), \; \Psi_{s}(x) =
 \sum\limits_{n} \Psi _{s|n} \exp (iq_{n} x ).
\end{equation}
Here we introduce the dimensionless variable
\[
a_s= -i k \lambda_{ab}^2 p_s.
\]
Substituting the expressions (\ref{26_02_08})--(\ref{magsup}) in
Eqs.~(\ref{e88})--(\ref{granysl2}) gives a set of linear equations
which allows us to find the coefficients $\bar{\Psi}_{s|n}$,
$\Psi_{s|n}$ in the expansions \eqref{magsup} and the
eigen-numbers $p_s$. After excluding the coefficients
$\bar{\Psi}_{s|n}$, we arrive at the set of equations for $\Psi
_{s|n}$. It can be solved by perturbations with respect to the
small modulation,  $|f_{n}| \ll 1$. In linear approximation, and
in the absence of the degeneracy of the corresponding matrix,
i.e., at
\begin{equation}\label{26_02_08_1}
q_s^2 \ne q_n^2 \q  {\rm for } \q s\ne n ,
\end{equation}
we obtain
\begin{eqnarray}\label{eigen}
\nonumber &\displaystyle\Psi _{s|n}  = \delta _{s,n}  + \tilde
\Psi _{s|n}, \;\; \displaystyle\tilde \Psi _{s|n}  \simeq
\displaystyle \frac{q_{s}^2 }{q_{n}^2  -
q_{s}^2}  \tau_{n-s} , \;\; s \ne n , \\
&\tau_{s} = \displaystyle \frac{ f_{s } }{1 - \Omega ^2  - i\nu \,\Omega } , \q |\tau_{s} | \ll 1;
\end{eqnarray}
\[
p_s  \simeq \displaystyle\frac{1}{{\lambda _{ab} }}\sqrt {1 +
\displaystyle\frac{{\lambda _c ^2 q_{s}^2 }} {{1 - \Omega ^2  -
i\nu \Omega }}}  + O(|\tau|^2 ),
\]
\begin{equation}\label{eigen1}
\displaystyle{\mathop{\rm Re}\nolimits} [p_s ] > 0,\quad
\displaystyle{\mathop{\rm Im}\nolimits} [p_s ] > 0 ,
\end{equation}
where $\delta _{s,n}$ is the Kronecker delta,
$\Omega=\omega/\omega\!_{_J}$, $\nu=\omega_r/\omega\!_{_J}$.

Matching the tangential components of the electric and magnetic
fields at the interface $z=0$, we obtain an infinite set of linear
algebraic equations for the coefficients $C_s$ and their relations
to the TCs $R_n$:
\begin{equation}\label{dns}
\sum\limits_s {D_{n|s} C_s }  = 2 \beta^V \delta_{n,0} ,
\end{equation}
\begin{equation}\label{rn}
R_n  = \sum\limits_s {C_s \Psi _{s|n} }  - \delta _{n,0},
\end{equation}
where
\begin{eqnarray}\label{infset}
 &D_{ n|s}  = b_n \delta _{n,s}  + d_{n|s} ,\quad b_n  = \beta_n^V  + a_n , \\
 &d_{n|s}  = (\beta_n^V  +a_s )\tilde \Psi_{s|n}.
\end{eqnarray}

To solve the infinite set of equations~(\ref{dns}) for $C_s$ we
use resonant perturbation theory, which allows presenting results
in an explicit analytical form~\cite{katsetc}.

When all spatial field harmonics are far away from the eigen-modes
of the unmodulated layered superconductor (nonresonance
conditions), the diagonal elements $b_{s}$ of the matrix $\hat{D}
\eq \| D_{n|s}\|$ are of the order of one or larger,
$|b_{s}|\sim|\beta_{s}^V|\gtrsim 1$. In this case, the matrix
$\hat{D}$ is diagonal-dominated, that is, its off-diagonal
elements are small compared to the diagonal ones, $|d_{n|s}| \sim
|\tau_{n-s}| \ll |b_s|$. Then, the solution of Eqs.~\eqref{dns},
\eqref{rn} gives us a trivial result: the specular reflection TC,
$R_0$, is close to the Fresnel coefficient,
\begin{equation}\label{5_03_08_3A}
    R_F  = \frac{\cos\theta  - a_0 }{\cos\theta  + a_0} \eq |R_F| \exp(i\psi),
\end{equation}
and differs from it by terms proportional to $\tau ^2$. Other TCs
are small, $R_n \sim \tau_n$, $n\ne 0$.

A much more interesting case occurs under the resonance
conditions, when \equ{1} holds for one (or simultaneously for two)
spatial field (resonance) harmonics,
\begin{equation}\label{25_12_07}
q_r = k\sin\theta + r g \simeq \mathrm{sign} (r)
\mathrm{Re}[\kappa_{\rm sw}] .
\end{equation}
Here $r> 0$ ($r<0$) corresponds to  the forward (backward)
propagation of the excited SJPW  with respect to the incident
wave.

For simplicity, we restrict ourselves to the single-resonance
case. In the resonant case, the diagonal matrix element $D_{r|r} =
b_r$ becomes anomalously small, and the determinant of the matrix
$\hat{D}$ decreases significantly (see, e.g.,
Ref.~\onlinecite{katsetc}). Recall that the normalized
$z$-component of the wave-vectors in vacuum, $\beta_{s}^V$, can be
either purely real or purely imaginary. Therefore, the minimum of
$|b_{r}|\ll 1$ holds in the vicinity of the point in the
$(\omega,\theta)$-plane where $\mathrm{Im}[\beta_{r}^V]=-
\mathrm{Im}\left[a_{r}\right]$, which is the dispersion relation
for SJPWs, \equ{25_12_07_1}.

Thus, the set of equations  \equ{dns} consists of one resonance
equation (with $n=r$),
\begin{equation}\label{27-02-08}
D_{r|r} B_{r}  + \sum\limits_{N \ne r} D_{r|N} B_{N} = 0,
\end{equation}
and the subset of nonresonance equations  (with ``nonresonance
numbers'' $N\ne r$). Solving the subset for the nonresonance
coefficients $B_{N}$ we obtain
\begin{equation}\label{g18}
B_{N}= 2\beta^V (\hat M^{ - 1})_{N|0}  -  B_{r}
\sum_{N'}(\hat M^{ - 1})_{N|N'} D_{N'|r} ,
\end{equation}
where $\hat{M}^{-1} $ is the matrix inverse of the nonresonance
square submatrix  $\hat{M} =\|D_{N|N'} \|$. Substituting $B_{N}$
in \equ{27-02-08} we obtain
\begin{equation}\label{5_03_08}
B_{r} = \frac{F_r}{\tilde{D}_{r|r} } ,
\end{equation}
where
\begin{equation}\label{5_03_08_1}
\tilde{D}_{r|r} = {D}_{r|r} - \sum_{N,N'}D_{r|N} (\hat M^{ - 1})_{N|N'} D_{N'|r},
\end{equation}
\begin{equation}\label{5_03_08_2}
F_r =- 2\beta^V \sum_{N} D_{r|N}(\hat M^{ - 1})_{N|0} .
\end{equation}

We now examine the solution Eqs.~\eqref{g18}, \eqref{5_03_08} in
the main approximation, i.e., taking into account the
linear-in-$\tau$ term in $F_{r}$,
\begin{equation}\label{5_03_08_4}
F_r =- \frac{ 2\beta^V d_{r|0}}{b_0 } ,
\end{equation}
and quadratic-in-$\tau$ terms in $\tilde{D}_{r|r}$,
  \begin{equation}\label{5_03_08_5}
\tilde{D}_{r|r} = \beta_{r}^V  + a_r + C_{r} , \q C_{r} = -
\sum\limits_N \frac{d_{r|N} d_{N|r}}{b_N} .
\end{equation}
In this approximation we keep only the zero-order term in the
series expansion of $\hat{M}^{-1} \simeq \| \delta_{N,N'}/b_N\| $.
Thus, we obtain
\begin{equation}\label{crcn}
B_{r}  = \frac{F_r}{\beta_{r}^V  + \xi \beta_r + C_{r} },
\end{equation}
\begin{equation}\label{5_03_08_3}
B_{N}  = \frac{2 \beta^V \delta_{N,0} -
d_{N|r} B_{r} }{b_N } .
\end{equation}

Finally, using Eqs.~(\ref{rn}), (\ref{eigen}) we derive the
resonance, $R_r$, and nonresonance, $R_N$, transformation
coefficients,
\begin{equation}\label{rrrn}
R_r= B_{r} ,\;\; R_N  = R_F \delta _{N,\,0}  +R_r
\left( \tilde{\Psi} _{r|N} - \frac{d_{N|r}}{b_N }\right).
\end{equation}
It is convenient to present the resonance TC, $R_r$, in the form
\begin{equation}\label{13_12_07_1}
R_r = \frac{F_{r}}{\beta_{r}^V  + a_r +C_r(\theta, \Omega, \tau)} ,
\end{equation}
where $C_r \eq C_r(\theta, \Omega, \tau)$ is the parameter that
describes the \emph{coupling} between waves in the vacuum and the
layered superconductor. Below we assume the coupling parameter
$C_r$ to be small. However, even when $|C_r| \ll 1$, the coupling
of the waves in the vacuum and superconductor plays a very
important role in the excitation of SJPWs and in the anomalies of
the reflection properties (Wood's anomalies).

First, the dispersion relation of the surface Josephson plasma
waves is modified, involving the radiation leakage in the vacuum.
The new spectrum of the SJPWs is defined by equating the
denominator in Eq.~(\ref{13_12_07_1}) to zero. Thus, the quadratic
in the modulation term, $C_{r}$, is responsible for the shift of
the position of the resonance, $\mathrm{Im}[C_{r}]$, and its
widening, $\mathrm{Re}[C_{r}]$. The region where the coupling
$|C_r|\ll 1$ (when the radiation leakage of the excited SJPW does
not dominate) corresponds to the strongest excitation of the
surface waves by the incident waves.

Second, due to the coupling, the specular reflection coefficient,
$R_0$, in Eq.~(\ref{rrrn}) differs from the Fresnel coefficient,
$R_F$, and its modulus becomes less than one. Moreover, as we show
below, the reflection of waves with any given frequency
$\omega<\omega_{_J}$ can be totally suppressed, for the specific
incident angle $\theta$ and the modulation magnitude. This
provides a way to control and filter the THz radiation.

In the next section, we study in detail the strong effects in the
excitation of the SJPWs, the enhancement of absorptivity, and the
suppression of the specular reflectivity near the resonance.

\section{Suppression of the specular reflection}

The transformation coefficient $R_0$ for the specularly-reflected
wave, \equ{rrrn}, can be rearranged as
\begin{equation}\label{r0}
R_0 =R_F \frac{k_{zr}^V/k  +a_r + C_r(\theta, \Omega,
\tau)-\Delta_r(\theta, \Omega, \tau)}{k_{zr}^V/k +a_r + C_r(\theta,
\Omega, \tau)},
\end{equation}
where $\tau$ stands for $\tau_r$, and
\begin{equation}\label{r01_OLD}
\Delta_r(\theta, \Omega, \tau)=\frac{2
\cos\theta}{\cos^2\theta-a_0^2}\left(a_0-a_r\right)\tilde{\Psi}_{0r}\tilde{\Psi}_{r0}.
\end{equation}

To study the resonance phenomena, we consider the case most
suitable for their observation, when the following inequalities
are satisfied:
\begin{equation}\label{27_02_07A}
\frac{D^2}{\lambda_{ab}^2} \sin^2\theta \ll
(1-\Omega^2)\varepsilon \ll 1 .
\end{equation}
The left inequality corresponds to the continuum limit for the
field distribution in the $z$-direction, whereas the right
inequality allows neglecting unity under the square root in
Eq.~(\ref{eigen1}). Besides, we assume the dissipation parameter
$\nu$ to be small as compared to $(1-\Omega^2)$,
\begin{equation}\label{28_02_07}
\nu\ll \left(1-\Omega^2\right).
\end{equation}
For this frequency region, the complex parameter
$a_r=a_r(\theta,\Omega) \eq a_r^\prime+ia_r^{\prime\prime}$ can be presented as
\begin{equation}\label{27_02_07-2}
a_r=\frac{k^2
  \lambda_{ab}\lambda_c}{2\sqrt{1-\Omega^2}}
  \left(\frac{\nu}{1-\Omega^2}-2i\right)|\bar{q}_r|,
\end{equation}
where we introduce the dimensionless variable
\[\bar{q}_n=\frac{{q}_n}{k}=\sin\theta+n \frac{g}{k}.\]

When restrictions Eqs.~(\ref{27_02_07A}), (\ref{28_02_07}) are
valid, the expression for the reflectivity coefficient can be
significantly simplified. First, the phase $\psi$ of the Fresnel
reflectivity coefficient, Eq.~(\ref{5_03_08_3A}), is small,
\begin{equation}\label{28_02_07-1}
\psi\simeq 2\frac{k^2
\lambda_{ab}\lambda_c}{\sqrt{1-\Omega^2}}\tan\theta \ll 1.
\end{equation}
Second, the parameter $a_r$ in Eq.~(\ref{27_02_07-2}) depends
weakly on the angle $\theta$ in the vicinity of the resonance,
whereas it depends strongly on the frequency detuning
$(1-\Omega)$, and its real part is sensitive to the magnitude
$\nu$ of the damping. Note also that near the resonance,
$\Delta_r(\theta,\Omega,\tau)$ in Eq.~(\ref{r0}) is almost real,
$\Delta_r(\theta,\Omega,\tau) \simeq - 2 {\rm
Re}\,[C_r(\theta,\Omega,\tau)]$.

Vanishing the imaginary part of the denominator in
Eqs.~\eqref{13_12_07_1}, \eqref{r0},
\begin{equation}\label{10_08C1}
{\rm Im}[k_{zr}^V/k  +a_r + C_r(\theta, \Omega, \tau)]=0,
\end{equation}
defines a curve in the $(\Omega,\theta)$-plane, where
$|R_r(\Omega,\theta)|$ achieves its maximum. In view of assumed
smallness of the coupling coefficient $C_r$, this curve passes
close to
\begin{equation}\label{10_08D}
\theta =\theta_0 \eq \arcsin\left|1-r\frac{g}{k}\right| .
\end{equation}

Separating the real and imaginary parts in the numerator and
denominator in Eq.~(\ref{r0}), we rewrite the specular reflection
coefficient $R_0$ the form,
\begin{equation}\label{28_02_07-4}
R_0=\frac{X_r(\vartheta,\Omega)+i\Bigl[{\rm Re}[C_r(\Omega,
\tau)]-C_{\rm opt}(\Omega)\Bigr]}{X_r(\vartheta,\Omega
)-i\Bigl[{\rm Re}[{C}_r(\Omega, \tau)]+C_{\rm opt}(\Omega)\Bigr]},
\end{equation}
where we introduce the incident-angle deviation $\vartheta =
\theta- \theta_0$. For  simplicity, below  we restrict ourselves
to the case of harmonic modulation and consider the resonances in
the plus- and minus-first orders, $r=\pm 1$. Then,
\begin{equation}\label{10_08G}
X_r(\vartheta ,\Omega ) \!\simeq \! r\cos\theta_0\frac{{1 - \Omega ^2 }}{{k^4
\lambda _{ab}^2 \lambda _c^2 }}\cdot \left(\vartheta  - \vartheta_{\rm res}\right) ,
\end{equation}
\begin{equation}\label{28_02_07-8}
\vartheta_{\rm res}  \simeq 4\frac{{2 - g^2 /k^2 }}{{(4 -
g^2 /k^2 )^2 }}\frac{k^4 \lambda _{ab}^2 \lambda
_c^2}{1-\Omega^2}\,\, \frac{|\tau_r |^2}{\cos\theta_0} ,
\end{equation}
\begin{eqnarray}\label{10_08j}
{ \rm Re}[C_r(\Omega ,\tau )] \simeq\,
 \sqrt{\frac{k}{g}} \frac{(1-rg/k)^2}{(2-rg/k)^{5/2}}\,  \frac{{k^2 \lambda _{ab} \lambda
_c }}{{\sqrt {1 - \Omega ^2 }
 }} \, |\tau _r |^2,
\end{eqnarray}
\begin{equation}\label{10_08F}
C_{\rm opt}(\Omega)\simeq \frac{\nu }{{2(1 - \Omega ^2) }}.
\end{equation}
Equations~(\ref{28_02_07-4})--(\ref{10_08F}) show that the modulus
of the specular reflectivity $R_0(\theta)$ has a sharp resonance
minima at $\vartheta =\vartheta_{\rm res}$,
\begin{equation}\label{28_02_07-9}
|R_0|_{\rm min}\simeq \frac{|C_{\rm opt}(\Omega)-{\rm
Re}[C_r(\Omega,\tau)]|}{C_{\rm opt}(\Omega)+{\rm Re}[C_r(\Omega,
\tau)]}.
\end{equation}
Its angular width, $\delta\vartheta$, is
\begin{equation}\label{28_04_08}
\delta\vartheta = \frac{\nu}{(1-\Omega^2)^2} \frac{k^4\lambda_{ab}^2
\lambda_{c}^2}{\cos\theta_0} \ll 1 .
\end{equation}

It is clearly seen that $|R_0|_{\rm min}$ depends strongly on the
frequency detuning $(1-\Omega)$, dissipation parameter $\nu$, and
the coupling between waves in the vacuum and the layered
superconductor, i.e., on the modulation magnitude $|f_r|$. This
offers several important applications of the predicted anomaly of
the reflectivity in the THz range. For instance, if the coupling
parameter ${\rm Re}[C_r(\Omega,\tau)]$ is equal to  $C_{\rm opt}$,
i.e., the modulation magnitude $|f_r|$ takes on the optimal value,
\begin{equation}\label{28_02_07-10}
|f_r|_{\rm opt}^2\simeq
\frac{\nu}{2}\sqrt{\frac{g}{k}}\frac{(2-rg/k)^{5/2}}{(1-rg/k)^2}
\frac{(1-\Omega^2)^{3/2}}{k^2\lambda_{ab}\lambda_{c}},
\end{equation}
then the specular reflection coefficient $R_0$ at
$\vartheta=\vartheta_{\rm res}$ vanishes. This means that, by
appropriate choice of the parameters, the total suppression of the
reflectivity can be achieved due to the resonance excitation of
the surface Josephson plasma wave.

In the vicinity of the resonance, the relative amplitude of the
excited SJPW can be approximated by
\begin{equation}\label{28_04_08_1}
R_r \simeq 2i \frac{ \tau_r \c
(1-\sin\theta_0)\tan^2\theta_0}{X_r(\vartheta,\Omega )-i\Bigl[{\rm
Re}[{C}_r(\Omega, \tau)]+C_{\rm opt}(\Omega)\Bigr]} .
\end{equation}
Note that equations $\vartheta=\vartheta_{\rm res}$ and ${\rm
Re}[C_r(\Omega,\tau)] =C_{\rm opt}$ (i.e., $|f_r| =|f_r|_{\rm
opt}$) constitute the conditions not only for the total
suppression of the specular reflection, but also for the best
matching of the incident wave and the SJPW. Under such conditions,
the amplitude of the excited surface wave is much higher than the
amplitude of the incident wave,
\begin{equation}\label{01_03_07}
|R_r|_{\rm max}\simeq\frac{\sqrt{2}(1-\Omega^2)^{3/4}} {k\sqrt{\nu
\lambda_{ab}\lambda_{c}}}\frac{\sin\theta_0}{\cos^{3/4}\theta_0}\gg
1.
\end{equation}
Thus, we can achieve a high concentration of THz radiation energy
in the SJPW.

The resonant decrease of the amplitude of the specularly-reflected
wave is accompanied  by the resonant increase of the absorption.
Evidently, for the optimal conditions, $\vartheta=\vartheta_{\rm
res}$, $|f_r| =|f_r|_{\rm opt}$, which correspond to the total
suppression of the specular reflectivity, the energy pumped into
the layered superconductor from the vacuum can be completely
transformed into Joule heat due to the quasiparticle resistance.
For the diffraction on the harmonic grating, the dependence of the
absorptivity coefficient $A$ on the wave frequency and the
incident angle is described by a resonance curve,
\[
A(\vartheta,\Omega)=1-|R_0(\vartheta,\Omega)|^2
\]
\begin{equation}\label{02_03_07}
\simeq \frac{4 C_{\rm opt}(\Omega){\rm Re}[{C}_r(\Omega,
\tau)]}{X_{r}^{2}(\vartheta,\Omega )+\bigl[{\rm Re}[{C}_r(\Omega,
\tau)]+C_{\rm opt}(\Omega)\bigr]^2} ,
\end{equation}
accurate within terms of order $|\tau_r|^2$. It should be noted
that the resonance increase of the electromagnetic absorption can
result in a transition of the superconductor into normal state.
Thus, new kinds of resonance phenomena can be observed in layered
superconductors due to the excitation of the SJPWs. However, for
rather low intensities of the incident wave, the sample heating
due to the Joule losses can be neglected.

\begin{figure}
\begin{center}
\begin{center}
\vspace*{-17cm}
\hspace*{2cm}\includegraphics[width=25cm]{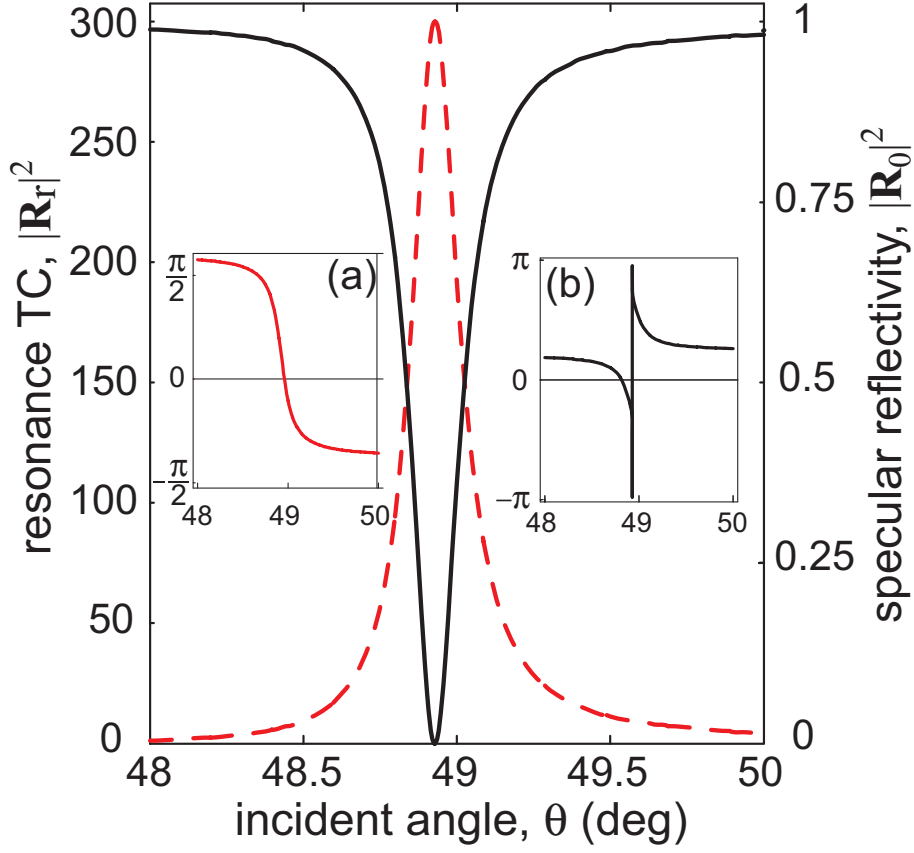}
\end{center}
\caption{(Color online) Numerical simulation of the total
suppression of the specular reflection. Black solid and red dashed
curves show the angular dependences of the specular and resonance
TCs, respectively, for the forward resonance diffraction ($r = 1$)
on the harmonic grating. These calculations were performed using
Eqs.~(\ref{13_12_07_1}), (\ref{r0}) for the harmonic grating with
pitch  $L$ = $1$~mm  and modulation magnitude $\left| {f_r
}\right| = 3.65\c 10^{-6}$. Other parameters used here are: $D =
2\cdot 10^{ - 7} $~cm, $\lambda _{ab} = 2.5 \cdot 10^{ - 5} $~cm,
$\nu  = 10^{ - 7} $, $\varepsilon = 20$,  and $(1-\Omega^2) =
1.2\cdot10^{-5}$. The insets show the angular dependences of the
phases of the resonance (a) and specular (b) TCs. } \label{graph2}
\end{center}
\end{figure}

We have illustrated our analytical results by the numerical
calculations of the specular and resonance TCs given in
Eqs.~\eqref{r0}, \eqref{13_12_07_1}. The angular dependences of
$|R_0|^2$ and $|R_r|^2$  for the forward resonance diffraction in
the first diffraction order ($r=+1$) on a harmonic grating are
shown in Fig.~\ref{graph2}. The asymptotic
formulae~(\ref{28_02_07-4}), (\ref{28_04_08_1}) are in a good
agreement with these plots. The modulation magnitude, $|f_r|$, was
chosen to achieve the total suppression of the specular
reflection. Its value is close to  $|f_r|_{\rm opt}$ defined by
the asymptotic expression (\ref{28_02_07-10}).

The minimum in the specular reflectivity (see Fig.~\ref{graph2})
is caused by the destructive interference of the waves scattered
via two different channels. The first channel is the direct
(approximately total) reflection from the unmodulated
vacuum-layered superconductor interface. The magnetic-field
amplitude of this wave is approximately equal to that of the
incident wave, $R_F \simeq 1$. The second channel is defined by a
two-step scattering process: the diffraction of the incident wave
into the SJPW and the re-scattering of the SJPW into the specular
direction. By means of Eqs.~\eqref{28_04_08_1},
\eqref{28_02_07-4}, one can easily follow the phase changes in the
resonance and specular TCs. The value of $X_r(\vartheta, \Omega)$
changes its sign when $\vartheta$ crosses the point
$\vartheta_{\rm res}$. Correspondingly, the $R_r$ experiences the
phase shift $\sim \pi$ while the phase of $R_0$ changes by $\sim 2
\pi$ (see insets in Fig.~\ref{graph2}).

\begin{figure}[hbpt]
\vspace*{-22cm} \hspace*{2cm}
\includegraphics[width=25cm]{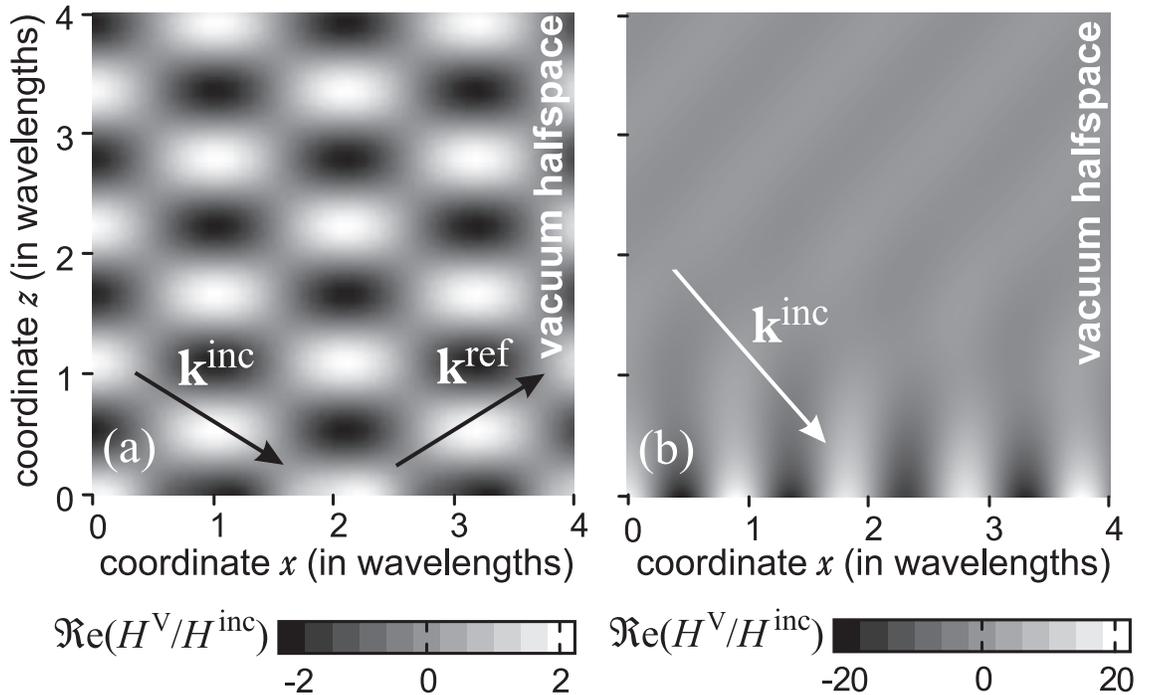}
\caption{The magnetic field distribution for the non-resonant
case, $\theta =60^o $, shown in (a), and for the resonant
diffraction in diffraction order $+1$, $\theta=\theta_{\rm
res}=48.93^o$, shown in (b). Other parameters are the same as in
Fig.~\ref{graph2}.}\label{f3}
\end{figure}

We also illustrated the effect of total suppression of the
specular reflection by the distribution of the total magnetic
field in the vacuum, Fig.~\ref{f3}. The interference pattern is
seen for the non-resonant case, when the amplitudes of the
incident and reflected waves practically coincide. Under the
resonance condition, when the reflected wave is totally
suppressed, the interference pattern in the far field disappears,
while the near-field ``torch'' structure of the SJPW is clearly
seen near the vacuum-layered superconductor interface.

\section*{Conclusion}\label{Conclusion}

In this paper, we present the detailed examination of the
resonance features for the diffraction of THz radiation on
periodically-modulated layered superconductors. The resonance is
caused by the excitation of the SJPWs for definite combinations of
the incident angle and frequency, and is analog to the widely
studied \emph{surface plasmon polariton resonance} in the visible
and near-infrared region. The analytical approach developed here
allows us to predict strong resonance effects (total suppression
of the specular reflection and total absorption) for specific
combinations of the parameters. The simplest (in-plane)
configuration for $TM$-polarized incident wave was examined here
under single-resonance conditions (i.e., excitation of one running
SJPW). This approach allows a similar study of the simultaneous
excitation of two SJPWs (double resonance), as well as the
examination of the so-called ``conical diffraction mount''
(out-of-plane diffraction). These items will be studied in the
future. It seems interesting also to consider the resonance
diffraction features for superconducting films of finite
thickness. There, the effects of resonance enhancement of the
transmissivity could exist.

The strongly selective interaction of SJPWs with the incident wave
having a certain frequency and direction of propagation can be
used for designing future THz detectors and filters. For instance,
the simplest design of a THz detector could be built around a $\rm
BiSrCuCaO$ sample fixed on a precisely rotated holder and attached
by contacts to measure its resistance. When rotating the sample,
the incident THz radiation would produce a surface wave at certain
angles. This results in a strong enhancement of the absorption.
Respectively, the sample temperature increases, thus, its
resistance would increase.

\begin{acknowledgments}

We gratefully acknowledge partial support from the National
Security Agency (NSA), Laboratory of Physical Sciences (LPS), Army
Research Office (ARO), National Science Foundation (NSF) grant No.
EIA-0130383, JSPS-RFBR 06-02-91200, and Core-to-Core (CTC) program
supported by Japan Society for Promotion of Science (JSPS). S.S.
acknowledges support from the Ministry of Science, Culture and
Sport of Japan via the Grant-in Aid for Young Scientists No
18740224, the EPSRC via No. EP/D072581/1, EP/F005482/1, and ESF
network-programme ``Arrays of Quantum Dots and Josephson
Junctions''.

\end{acknowledgments}

\end{document}